# Exploring Tenets of Data Democratization

*Emergent Research Forum (ERF)*

**Sasari S. U. Samarasinghe**
University of Southern Queensland
Springfield, Australia
udanjala@gmail.com

**Sachithra Lokuge**
University of Southern Queensland
Springfield, Australia
ksplokuge@gmail.com

**Lan Snell**
University of Southern Queensland
Toowoomba, Australia
lan.snell@usq.edu.au

## Abstract

Data democratization is an ongoing process that broadens access to data and facilitates employees to find, access, self-analyze, and share data without additional support. This data access management process enables organizations to make informed decisions, which in return enhances organizational performance. Technological advancements and extensive market pressure have mandated organizations to transform their traditional businesses into data-driven organizations, focusing on data democratization as a part of their data governance strategy. This paper explores the tenets of data democratization through an in-depth review of the literature. The analysis identified twelve attributes that enable data democratization based on the literature review. Future work will focus on testing and further empirically investigating these to develop a framework for the data democratization process to overcome the challenges.

## Introduction

Data is a vital asset for any organization. The changing data architecture and the growing involvement of employees with data have inevitably introduced new and efficient processes such as data democratization (Gibson 2022). Democratization is, by definition, an ongoing process which has been conceptualized in myriad ways, such as a demand, a set of institutional changes or a form of political domination with an extension of citizenship rights and relates to political propaganda frequently (Grugel and Bishop 2013). Data democratization refers to a process of providing broader access to organizational data for both technical and non-technical users (Awasthi and George 2020). It has evolved from the concept of "democratization" (Ramamurthy 2019). It is also defined as an enterprise capability that motivates and empowers a broader scope of employees within the organization to find, access, interact, and share data in a secure and acquiescent way (Lefebvre et al. 2021). Data democratization helps organizations to convert data into a more imputable form for all employees who need access to data for problem-solving and decision-making (Shamim et al. 2021). Further, it represents the organizational understanding of FAIR principles (i.e., findable, accessible, interoperable, reusable) that encourage employees to unlock value from data (Labadie et al. 2020). When organizational data is democratized and backed by a supportive culture, it promotes knowledge sharing and willingness to accept diversity in new knowledge that traditional job functions of employees did not realize (Hyun et al. 2019). However, the end goals of data democratization are to empower employees (Hertzano and Mahurkar 2022), promote accurate decision-making, and ultimately gain a competitive advantage (Awasthi and George 2020). Therefore, with the establishment of data-driven organizations and modern business challenges, organizations moving toward data democratization requires identifying the attributes of data democratization, which will help to form their data and governance strategy accurately.

The advent of digital technologies and external triggers such as the COVID-19 pandemic has exerted additional pressures on organizations to digitize their business data (Lokuge and Sedera 2020). As a result, contemporary organizations foster plenty of upsides and growing interest in data democratization more than ever. In such a setup, quick and faster access to data and decision-making are vital for the growth of the business. However, while the concept of data democratization paints a positive picture since most users who benefit and gain access to data through data democratization are non-technical users, the results may be less accurate due to the lack of data science knowledge (Labadie et al. 2020). Besides, skills gaps and data awareness gaps are significant challenges organizations face with implementing data democratization (Awasthi and George 2020). To overcome these obstacles, organizations should develop and execute a data democratization process that allows access to the right set of employees to support establishing a hyper-collaborative work culture and knowledge development (Hyun et al. 2019). Therefore, this research focuses on identifying the attributes of data democratization in a more profound approach, which organizations can refer to when planning their data democratization process. It also contributes to the existing knowledge on data democratization in information systems (IS) literature. To investigate this phenomenon, the overarching research question developed for this study is: *"What are the important attributes of data democratization?"*

The structure of the paper is as follows. The next section provides the research methodology followed in this study. Next, the findings of the literature review are provided. The conclusion section highlights the future research agenda and limitations of the study.

## Research Methodology

An in-depth systematic literature review was conducted to investigate the concept of data democratization. Data democratization is an emerging topic, and therefore, an extensive literature review provides a profound understanding of existing studies related to the research phenomenon and identifies the research gaps (Tate et al. 2015). The literature review followed a comprehensive search strategy described as follows. First, a set of online databases were selected related to IS discipline. As per Charrois (2015), the rule of thumb for searching literature in a systematic review is to use more than two databases. This study used eight databases, namely AIS virtual library, Wiley online library, Taylor & Francis online library, Emerald insight, Springer, Science Direct, IEEE, and ACM Digital Library. Second, the keywords were determined related to the topic of investigation and used only English language papers for the analysis due to researchers' language fluency limitations. For the keywords search, authors used the keyword "data democratiz(s)ation" OR "democratiz(s)ation of data" OR "democratiz(s)ed data" in titles or abstracts. The search result included sixty-five (65) papers from various perspectives in multiple disciplines. The search result highlighted that most of the papers related to data democratization were published in the last decade. Therefore, the researchers used papers published in 10 years, from 2013-to 2022. We excluded papers that did not define the term data democratization or attributes from IS perspective and removed the papers that included data democratization only as a supportive term to their research study with no in-depth discussion. The researchers looked for terms such as "data democratiz(s)ation" OR "democratiz(s)ation of data" OR "democratiz(s)ed data." We found ten (10) papers that were directly relevant to "data democratiz(s)ation" and/or have defined and discussed "data democratiz(s)ation" related to IS discipline with the link to concepts such as "open data access to the public", "democratizing public and government data", "democratization of IT", and "role of democratization in data-driven organizations". Each paper was read entirely, analyzed, and categorized using essential information related to data democratization to identify the attributes in multiple disciplines.

## Analysis and Discussion

Democratization is often spanned into political and social sciences, with a broader definition of an ongoing process which has variously been conceptualized as a demand, a set of institutional changes or a form of political domination with the creation of an extension of citizenship rights to facilitate free and fair election (Grugel and Bishop 2013, p.5). It has derived from the word "democracy" which relates to politics than economics, but later the term has been widely used in technology (Karlovitz 2020), business (Bayer and Öniş 2010) and other industries gaining popularity among scholars. Democratization of technology refers to an ongoing process in which technology is rapidly becoming more accessible to more people (Clark et al. 2021). Democratization of information refers to providing extended power of using business information created by employees to all other employees, allowing them to access information, make better decisions, and improve visibility and transparency (Hollander 2019). Similarly, the democratization of data in organizations has gained popularity in solving data challenges in organizational operations.

The term 'Open Data,' defined as data that are free to use, reuse and redistribute (Doctor and Joshi 2021) - has often captured scholars' attention as it promotes the value of reusing data through active collaboration among citizens, governments and organizations (Espinosa et al. 2014). Some ideate that data democratization is similar to 'open data' concept, and many misconceptions regarding data democratization exist in IS research. As such, a deep dive into the conceptualization of data democratization is warranted. Most scholars have focused on open data and big data analytics. Scholars such as Shamim et al. (2021) introduced "Big Data Democratization", which refers to an organization's ability to convert big data into more referable language for all employees in need of solving business problems. Concerning the term 'democratization,' data democratization is considered a 'process'; rather than a destination (Hinds et al. 2021, p. 67). In an organizational setting, data should be democratized to facilitate the use of data analytics, which favorably improves organizational agility –how data informs decision making promptly - (Hyun et al. 2019, p. 3). However, due to the misuse of data and erroneous assumptions without proper training and knowledge among employees (Awasthi and George 2020), data democratization may hinder the value creation of democratizing inter-organizational business data.

**Defining Data Democratization**

Based on the sample, it was evident that scholars have a growing interest in studying data democratization, as many papers were published within the last three to four years. Table 1 presents the matrix of attributes related to the data democratization process from the analysis. Based on the findings in Table 1, it is evident that different authors have defined, and conceptualized data democratization based on various factors. As such, we propose data democratization as *"an ongoing process of enabling digital data access to both technical and non-technical users to understand, find, access, use, interact and share appropriate data within the boundaries of legal, confidentiality and security limitations by transferring data ownership and responsibility to empower users for efficient and accurate decision-making, promote collaboration, and create a knowledge-sharing culture in an organization."* The specification of the term 'legal confidentiality and security limitation' captures widely identified roadblocks in data democratization. When data is accessible to all users, the probability of data misuse can be high. Therefore, digital data access must be granted to each user based on their eligible job role within the organizational data hierarchy. Once data democratization is implemented, and users are well-trained to uplift their data literacy, it empowers and motivates them to use data accurately in their jobs. As such, it will positively influence their ability to make better decisions.

# Conclusion

The aim of this paper was to review and understand the literature on data democratization to identify the attributes of data democratization with future research ideas. A systematic review of 10 papers in 8 online databases was considered for this study. Based on the analysis, the authors identified 12 attributes of data democratization, critically analyzed the existing definitions of data democratization, and introduced a new definition that captures all-important attributes of data democratization. While we acknowledge the sample size is small, the authors of this paper have conducted an in-depth analysis to provide a clear overview of data democratization. Future research includes testing these in real-world scenarios to an in-depth understanding of data democratization and developing a framework. Various scholars conducted previous research studies to understand the relationship between data democratization to facilitate decision making and gaining competitive advantage. However, applying the knowledge-generating process through data democratization for employee learning and innovation is yet understudied. As such, future research can investigate the innovation aspect of data democratization, its implications, and enablers of data democratization from both organizational and individual perspectives.

| **Attributes of Data Democratization** | Hertzano and Mahurkar (2022) | Lefebvre et al. (2021) | Hinds et al. (2021) | Sitar-Tăut (2021) | Shamim et al. (2021) | Zotoo et al. (2021) | Awasthi and George (2020) | Hyun et al. (2020) | Labadie et al. (2020) | Hyun et al. (2019) |
|---|---|---|---|---|---|---|---|---|---|---|
| Broader data access to all users | | x | x | x | x | x | x | | x | |
| Transferring data ownership and responsibility to all users | | x | | | | | x | | | |
| Reasonable limitations on legal, security and data access rights | | | x | | | | x | | x | |
| Support problem solving and decision making | | | x | | x | x | x | x | | x |
| Achieve Competitive advantage and individual innovativeness through data value creation | | x | | | x | x | x | | | |
| Culture of willingness to share and accept diversity | | | | | x | | | x | | x |
| Remove obstacles to data exploration and sharing and preauthorization | x | | | | | x | | | x | |
| Easy to find, access, interoperable, and reuse data (FAIR) | | x | x | | | | x | | x | |
| Promote data sharing | x | x | x | | | | x | x | x | |
| Motivate and empower users | x | x | | | | | x | | x | |
| Use of analytics tools | | x | | | | | | x | | x |
| Collaboration and knowledge sharing (i.e., data & analytics skills) | x | x | | x | x | x | x | | | |

**Table 1. Attributes of Data Democratization**